\documentclass[aps,pra,twocolumn,superscriptaddress]{revtex4-2}
\usepackage{graphicx}
\usepackage{amssymb,amsfonts,amsmath}
\usepackage[colorlinks=true,citecolor=Cerulean,linkcolor=RubineRed,urlcolor=Cerulean]{hyperref}
\hypersetup{breaklinks=true}
\usepackage{graphicx}
\usepackage{color}
\usepackage[usenames,dvipsnames]{xcolor}
\usepackage{epstopdf}
\usepackage[normalem]{ulem}
\usepackage{bm}
\usepackage{bbold}
\usepackage{float}

\usepackage{amsmath}
\usepackage{amsfonts}
\usepackage{amssymb}
\usepackage{dsfont}
\usepackage{tikz}
\usetikzlibrary{trees}
\usepackage{svg}
\usepackage{booktabs}

\newcommand{\ket}[1]{|#1\rangle}

\begin{document}

%%%%%%%%%%%%%%%%%%%%%%%%%%%%%%%%%%%%%%%%%%%%%%%%%%%%%%%%%%%%%%%%%%%%%%%%%%%%%
%%%  Title and abstract
%%%%%%%%%%%%%%%%%%%%%%%%%%%%%%%%%%%%%%%%%%%%%%%%%%%%%%%%%%%%%%%%%%%%%%%%%%%%%

% \title{Non-local gates with neutral atoms in a thermal cavity}
\title{A thermal microwave bus for neutral atom quantum computing}
\author{Matthew J.H. Kendall}
\author{Christopher J. Watson}
\author{Michael Ben Shem}
\author{Jonathan D. Breeze}
\affiliation{\vspace{0.5em}Department of Physics and Astronomy, UCL, London WC1E 6BT, United Kingdom}
\
\date{\today}

\begin{abstract}
High-fidelity two-qubit gates in neutral-atom arrays rely on the Rydberg blockade, which is intrinsically short ranged and requires long range connectivity to be achieved through atom shuttling. We propose a four level architecture, where the typical ground state qubit can be leveraged for its long lifetime and the Rydberg states couple to a microwave cavity, allowing for long range cavity mediated gates. We first find the thermal dependence of two established protocols, the dispersive iSWAP native to the Tavis-Cummings model and the Controlled phase gate generated from driving the cavity. We simulate them under the presence of Rydberg decay, thermal photons, finite cavity linewidth and find fidelities which accompany closed form bounds. We then introduce the bichromatic Raman gate, which only virtually populates the Rydberg states and cancels dispersive shifts to mitigate both atomic and cavity decay, achieving a fidelity of $F = 0.997$. Finally, we consider a full optical tweezer array in a microwave cavity, and show that using a cavity mediated gate to close the periodic boundaries of the toric code shortens an error correction round by a factor of 2.3 for realistic array sizes. This was then applied to the wider family of Bivariate bicycle codes, and we find that it shortens a round of the gross code by a factor of 4.8.
\end{abstract}

\maketitle
\section{Introduction}
High-fidelity two-qubit gates are an essential requirement for fault-tolerant quantum computation. Neutral atoms in optical tweezers have emerged as a leading platform, with recent demonstrations of controlled-Z (CZ) gate fidelities exceeding $0.9985$ in $^{87}$Rb and $0.9959$ in $^{171}$Yb \cite{evered2026high, liu2026yb}. The typical gate mechanism uses the Rydberg blockade effect, in which two atoms shuttled to within each other's Rydberg blockade radius cannot be simultaneously excited, leading to a conditional phase \cite{Jaksch_2000}. The ability to shuttle qubits is an advantage of both neutral-atom and trapped-ion platforms, as it enables non-local connectivity that would otherwise require deep circuits \cite{Cohen2022,Bluvstein2022}. However, the time required to shuttle atoms across a planar array of length $L$ scales as $\mathcal{O}(L^{-1/2})$, which can become a bottleneck when implementing low overhead quantum error correcting codes such as quantum low density parity check (qLDPC) codes \cite{Breuckmann2021}. This motivates the investigation of entanglement bus architectures, in which quantum gates are implemented between distant atoms via an intermediary. This can be achieved by engineering the interactions to be longer range, such as using a chain of atoms to cascade the Rydberg blockade \cite{Delakouras2026}, or alternatively through the exchange of real or virtual photons between atoms coupled to a shared bosonic mode \cite{Duan2005}. The idea of using a common bosonic mode as an entangling bus originated from Cirac and Zoller \cite{CiracZoller}, who demonstrated that a shared phonon mode in a trapped-ion crystal could be used to implement a CZ gate. That protocol required the ions to occupy their motional ground state, a condition subsequently relaxed by the M{\o}lmer-S{\o}rensen (MS) gate \cite{S_rensen_1999}, which only virtually populates the mode, making the protocol robust to moderate thermal excitations and motional heating. The principles of the MS gate were then adapted to atoms coupled to a common photonic mode \cite{Guo}. In general, the gate infidelity scales as $\mathcal{O}(C^{-1/2})$, where $C = g^2/(\kappa\gamma)$ is the single-atom cooperativity, with $g$ the coupling strength and $\kappa$, $\gamma$ the cavity and atomic decay rates respectively \cite{Sorensen2003}. Notably, the cooperativity is independent of the magnitude of the atomic dipole moment, for a fixed wavelength, and is determined entirely by the ratio of the cavity quality factor $Q$ and the mode volume, and thus by the cavity geometry and composition alone. Cavity QED based MS gates have since been designed with more technical protocols, including the use of off resonant driving schemes \cite{Takahashi2017}, quantum Zeno generated blockade gates \cite{Srivastava2025}, and heralded approaches which reduces the infidelity scaling to $\mathcal{O}(C^{-1})$ \cite{Borregaard2015}. The cavity field can also be used to drive the gate, with no requirement for individual atom addressing and allowing for the implementation of arbitrary permutationally invariant unitaries \cite{Jandura2024, deliyannis2026}. Beyond individual gate protocols, cavity QED based architectures show promise in scalability, connectivity and for the efficient construction of non local entanglement resources for quantum error correction \cite{Ramette2022,interconnects, Chandra2025}.\\

Early experimental realisations began with the implementation of the protocol outlined in \cite{Guo} by Haroche using two flying Rydberg atoms in a microwave cavity \cite{haroche, arno}. Deterministic CZ gates were later realised in optical cavities with a fidelity of $0.76$, which was achieved with two rubidium atoms in a Fabry-Perot design \cite{Gerhard}. A source of infidelity in optical cavities is the sinusoidal spatial distribution of the mode, which causes $g$ to vary periodically along the cavity axis, imposing tight constraints on atom localisation at the field anti-nodes and thus requiring precise control of both the cavity lock and the tweezers. For applications in scalable quantum computing, where atoms must be positioned arbitrarily within a large array, this localisation requirement is a substantial practical challenge. Three-dimensional microwave cavities, however, have modes which are sufficiently large that fields can be engineered to be spatially uniform over the entire scale of the atomic register, eliminating the positioning requirement. Much of the extensive work on optical cavities for quantum information translates to microwave cavities, however a few considerations must be made. Firstly microwave frequencies correspond to transitions between Rydberg states, which have significantly shorter lifetimes. Secondly, the thermal photon population is non negligible at finite temperature. Therefore, gate protocols must be designed to be thermally robust through using interfering excitation pathways to cancel dispersive shifts \cite{Sarkarny_2015}. For planar approaches, such as an atom chip, the Rydberg transition couples to the evanescent field of the resonator and thermal effects can be mitigated through increased detuning from the resonator frequency  \cite{Tsiamis}. \\

In this work, we first consider iSWAP and CZ gates for neutral atoms coupled to a three dimensional microwave cavity at finite temperature. Our qubit is encoded in the ground state and for non local gates it is mapped to a Rydberg transition which is dispersively coupled to a microwave mode. The iSWAP gate naturally occurs in the single excitation subspace and the CZ gate can be generated by a cavity drive. We then introduce a bichromatic Raman gate, which generates a conditional phase through two detuned optical drives which create state dependant displacements on the cavity mode. The gate is designed for increased thermal robustness through detuning symmetry, and to mitigate decay through only virtually populating the Rydberg states. Finally, we look at the impact of the cavity mediated gates on quantum error correction codes with respect to shuttling approaches. We simulate the qubit rearrangement overhead required to implement the toric code and show that a hybrid scheme, using local Rydberg blockade gates for the bulk stabilisers and cavity-mediated gates for the non-local stabilisers at the periodic boundaries provides a speed-up for experimentally realisable parameters. We then extend this to the wider family of Bivariate bicycle codes.
 
\section{Cavity mediated Gates}
\begin{figure*}[htbp]
    \centering
    \includegraphics[width=18cm]{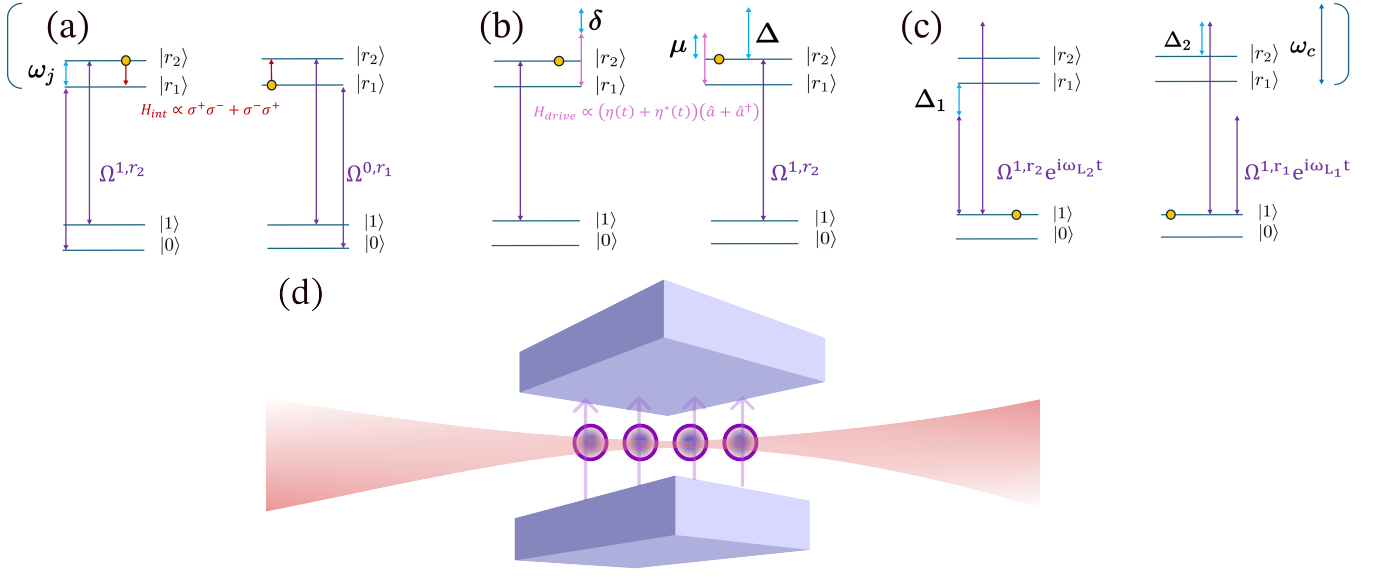}
    \caption{(a) Schematic of the iSWAP protocol. The ground state qubit
    $\{|0\rangle,|1\rangle\}$ are mapped to the Rydberg subspace
    $\{|r_1\rangle,|r_2\rangle\}$ via a resonant drive, during which the
    cavity-mediated exchange interaction drives the iSWAP operation, before
    the atoms are mapped back to the ground state. (b) Schematic of the geometric-phase CZ protocol. A $\pi$ pulse maps
    $|1\rangle\to|r_2\rangle$, leaving $|0\rangle$ dark and $|r_1\rangle$
    unpopulated, so the cavity-mediated exchange is absent. A drive $\eta(t)$ applied to the cavity displaces the mode along
    a state-dependent trajectory. When
    the loop closes at $t_g=2\pi\Delta/g^{2}$, the cavity disentangles and the
    enclosed area yields a geometric phase $\Theta\hat n^{2}$. (c) Schematic for the bichromatic Raman gate, two optical drives detuned from the $|1\rangle \rightarrow|r_1\rangle$ and $|1\rangle \rightarrow|r_2\rangle$ transitions with $\Delta_1$ and $\Delta_2$ respectively generate a state dependant force which completes a full cavity loop at $t_g = 2\pi /\delta_r$ where $\delta_r = \Delta_2-(\Delta_1+\Delta)$. (d) Schematic of the theoretical setup, a three dimensional microwave cavity with a tweezer array confined in the re entrant gap using optical tweezers.}
    \label{fig:schematics}
\end{figure*}
\subsection{Model}
In this work we consider an array of $^{87}$Rb atoms suspended in a 3D microwave cavity, held sufficiently far from the surfaces to minimise DC Stark shifts from stray electric fields and coupling to two-level systems at surfaces. The logical qubit \{$|0\rangle, |1\rangle$\} is typically stored in the hyperfine ground state of $5S_{1/2}$, which supports long coherence times and high fidelity single qubit gates. This is chosen to lie far below the cavity band, so the logical states are spectroscopically dark to the mode, the interaction is carried entirely by two auxiliary Rydberg states $\{|r_1\rangle,|r_2\rangle\}$ whose transition frequency $\omega_j$ is detuned from the cavity by $\Delta=\omega_j-\omega_c$. The cavity is engineered to host a mode with confined and homogeneous electric field, providing a region of strong and uniform coupling across the array. This is achievable in a re-entrant cylindrical geometry supporting a quasi-TM mode, in which the electric and magnetic field energies occupy spatially distinct regions and so resonant frequency and mode volume are set by the re-entrant gap \cite{leFloch2013}, cryogenic piezoelectric actuation of this gap provides in-situ tuning over GHz ranges \cite{Carvalho2016}. For frequencies $\omega \lesssim 2 \pi \times 30$ GHz with wavelengths larger than $1$ cm, through shrinking the gap between the posts to a few millimetres, we can achieve mode volumes as small as $V_{eff} \approx 10^{-3} \lambda^3$. The $50S_{1/2} \rightarrow 50P_{3/2}$ transition in $^{87}$Rb has a frequency $\omega \approx 2 \pi \times 30$ GHz and dipole moment $\mu \approx 1500 ea_0$. With $g=\mu E_{\rm vac}/\hbar$ and $E_{vac}=\sqrt{\hbar\omega_c/2\varepsilon_0 V_{\rm eff}}$, we obtain coupling $g$ on the order of $1$ MHz. Bulk niobium cavities prepared by optimised buffered chemical polishing reach single-photon internal quality factors above $10^9$ at 1.2~K \cite{oriani2025}, giving $\kappa/2\pi\approx20$~Hz and hence $g/\kappa\sim10^5-10^4$ and $C\sim10^6$--$10^7$ where we take the Rydberg decay rate as $\Gamma = 2\pi\times1$~kHz. Such $Q$ values were obtained in seamless coaxial geometries without optical ports, combining them with small $V_{eff}$ and the optical access required for tweezer trapping remains an engineering target, though seamless designs confining bound states below the waveguide cutoff show that optical accessibility need not preclude high $Q$. Finally, the thermal photon population follows $\bar n_{th}=[\exp(\hbar\omega_c/k_BT)-1]^{-1}$ is $2.3$ at 4~K and
$0.3$ at 1~K, falling below $10^{-4}$ for $T\lesssim100$~mK which means a cryogenic environment must be required to both reach superconductivity and reduce thermal photon population. Cryogenic atom arrays have been realised, a 4K heat shield extended the Rydberg lifetimes by over a factor of 3 reaching $\Gamma \approx 2\pi \times 1/3 $KHz, and shows promising scalability to thousand atom arrays \cite{cyro_atom, kumar2026, lim2026defect,zhang2025high, pichard2024rearrangement}. In this work we take $\omega_c = 2\pi \times 30$GHz, and $g = 2\pi \times 1MHz$. The respective detunings have been left as a free parameter, as both the cavity frequency and Rydberg transition frequency can be tuned. \\

The interaction of $N$ two-level atoms with a single cavity mode is described by the Tavis-Cummings Hamiltonian ($\hbar = 1$)
\begin{equation}
    H_{\mathrm{TC}} = \omega_c a^\dagger a + \frac{1}{2}\sum_{j=1}^{N} \omega_j \sigma_j^z
    + \sum_{j=1}^{N} g_j\!\left(\sigma_j^+ a + \sigma_j^- a^\dagger\right).
    \label{eq:TC}
\end{equation}
where $a$ ($a^\dagger$) is the cavity annihilation (creation) operator. In the dispersive regime $|\Delta_j| = |\omega_j - \omega_c| \gg g_j$, the cavity is only virtually populated and can be adiabatically eliminated via a Schrieffer-Wolff transformation (see Appendix A). To second order in $g_j/\Delta_j$, the effective Hamiltonian is
\begin{equation}
    \tilde H = \tilde{\omega}_c a^\dagger a
    + \frac{1}{2}\sum_{j=1}^{N} \tilde{\omega}_j \sigma_j^z
    + \sum_{j < k} J_{jk}\!\left(\sigma_j^+ \sigma_k^- + \sigma_j^- \sigma_k^+\right).
    \label{eq:Heff}
\end{equation}
where the cavity and qubit frequencies acquire dispersive shifts, and the cavity-mediated exchange coefficient is
\begin{equation}
    J_{jk} = \frac{g_j g_k}{2}\!\left(\frac{1}{\Delta_j} + \frac{1}{\Delta_k}\right).
    \label{eq:Jjk}
\end{equation}
The exchange interaction $\sigma_j^+\sigma_k^- + \sigma_j^- \sigma_k^+$ conserves total excitation number and therefore operates within the single-excitation subspace. Throughout this work we simulate this Hamiltonian as an open quantum system subject to the Lindblad collapse operators capturing cavity decay, $\hat L_{\kappa+} = \sqrt{\kappa (1+n_{th})}\hat a$, $\hat L_{\kappa-} = \sqrt{\kappa n_{th}}\hat a^\dagger$, where $n_{th}$ is the thermal cavity population. The Rydberg decay is then captured as $\hat L_{\Gamma_{r_i}} = \sqrt{\Gamma_i}|g\rangle\langle r_i|$ where $i = \{|r_1\rangle, |r_2\rangle \}$. We evaluate the average fidelity of the simulated logical channel $\mathcal{E}$ relative to the target unitary $U$. If $\{K_{\ell}\}$ denotes a Kraus representation of the projected logical
channel, the average fidelity is
\begin{equation}
F_{\mathrm{avg}}= \frac{ \sum_{\ell}
\left| \operatorname{Tr}
\left( U^{\dagger}K_{\ell}\right)\right|^{2}+
\sum_{\ell}
\operatorname{Tr}
\left(
K_{\ell}^{\dagger}K_{\ell}
\right)
}{
d(d+1)
}.
\label{eq:average-gate-fidelity}
\end{equation}
where $d=4$ for a two-qubit gate \cite{Pedersen_2007}.
\subsection{\label{sec:iswap}Dispersive iSWAP gate}
\begin{figure}
    \centering
    \includegraphics[width=8.4cm]{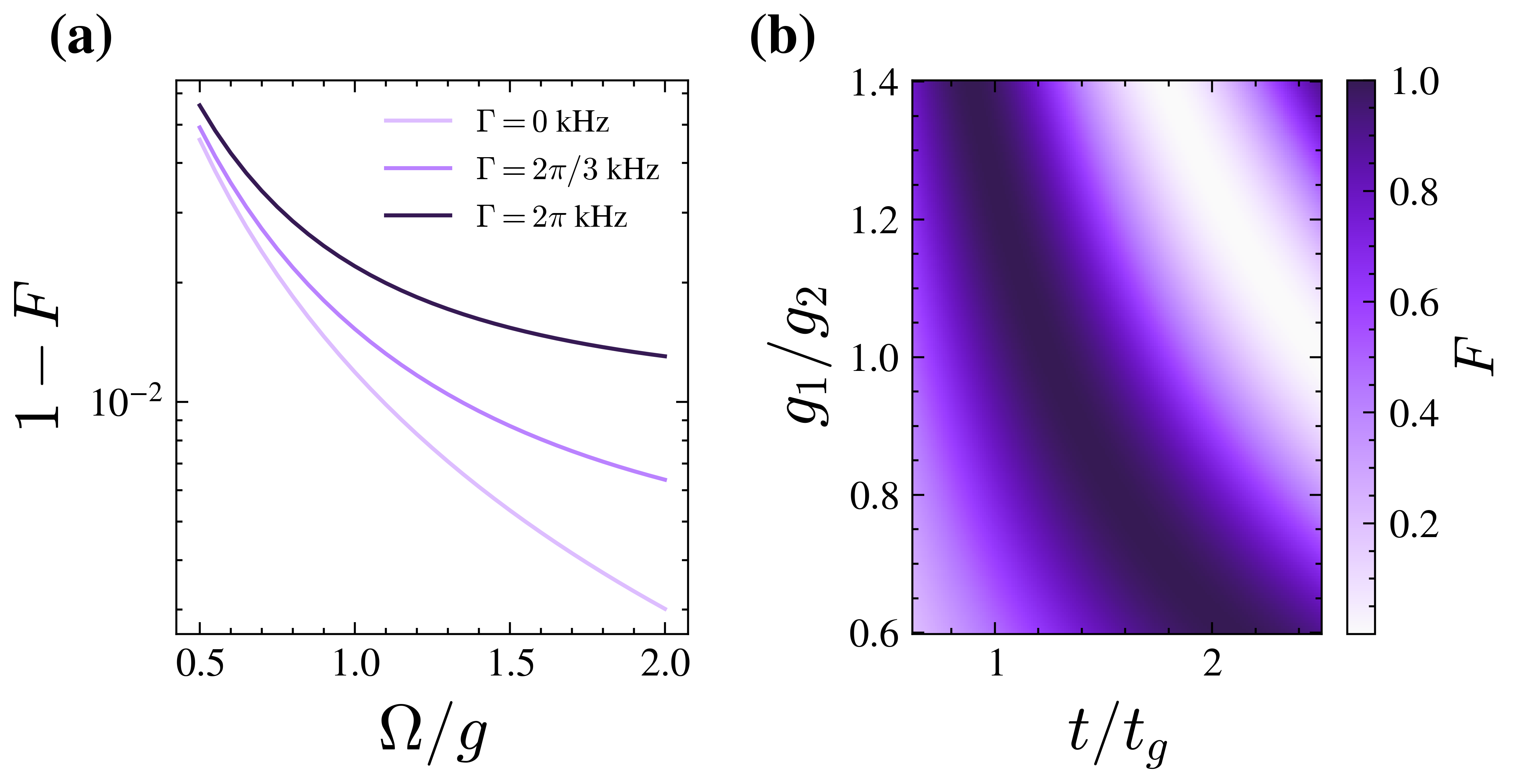}
    \caption{(a) Fidelity as a function of $\Omega/g$ and hence the time ratio  $T_{\mathrm{exc}}/T_{\mathrm{swap}}$. Simulated at $\Delta = 2\pi\times 20\mathrm{MHz}$. Fidelity falls for lower $\Omega/g$ as partial exchange happens during the excitation. (b) Coupling asymmetries can be corrected by tuning gate time or detuning.}
    \label{fig:fidelities}
\end{figure}

We now outline how to perform an iSWAP in the 4 level system, a nearest neighbour equivalent gate can be found in \cite{ildefonso2026expandingneutralatomgate}.
Under the exchange term in Eq. \eqref{eq:Heff}, the two-atom system undergoes population transfer at rate $J_{jk}$, realising an iSWAP gate after a half-period $T_{\mathrm{iSWAP}} = \pi/(2J_{jk})$. We first consider the symmetric case $g_{j,k} = g$ and $\Delta_{j,k} = \Delta$, where the swap rate simplifies to $J = g^2/\Delta = \chi$. The full protocol begins by coherently mapping the ground state qubit to the Rydberg states by the drive
\begin{equation}
\begin{split}
    H_{\mathrm{excite}}(t) = \sum_{j} \frac{1}{2}\!(
        \Omega_j^{0,r_1}(t)\,e^{i\phi_1}|r_1\rangle\langle 0|
        + \\
        \Omega_j^{1,r_2}(t)\,e^{i\phi_2}|r_2\rangle\langle 1|
        + \mathrm{H.c} ) .
\end{split}
\label{eq:Hdrive}
\end{equation}
where $\Omega_j^{0,r_1}$ and $\Omega_j^{1,r_2}$ are the Rabi frequencies coupling the ground states to the respective Rydberg levels and we take the time dependant pulse as a Gaussian envelope. Once the Rydberg manifold is populated, the exchange Hamiltonian natively mediates the exchange, after which the inverse drive maps the atoms back to the ground state. During the gate, the $| r_2r_2\rangle$ and $|r_1r_1\rangle$ states accumulate a photon-number-dependent dispersive phase relative to the single-excitation manifold, which is corrected through a global $| r_1\rangle\leftrightarrow| r_2\rangle$ echo. A schematic of the full protocol is shown in Fig. (\ref{fig:schematics})(a). Outside of decay, infidelity can come from the finite excitation time $T_{\mathrm{e}}$. Since the exchange Hamiltonian Eq.~\eqref{eq:Heff} is always present, partial iSWAP evolution occurs during the excitation pulse, leading to an error that grows inversely with the ratio $\Omega/g$, as shown in Fig.~(\ref{fig:fidelities}). The optimal point occurs at $\Omega/g = 2$ which corresponds to $F = 0.964$ and $F = 0.976$ at $\Gamma = 2\pi\times 1$kHz and $2\pi\times 1/3$kHz respectively. An inhomogeneous cavity field leads to site dependent couplings $g_j \neq g_k$, which can also degrade the gate. From Eq.~\eqref{eq:Jjk}, asymmetric couplings modify both the exchange rate and the dispersive shifts, with their effect on fidelity shown in Fig. (\ref{fig:fidelities})(b). The asymmetries can in principle be corrected for by restoring $J_{j,k}$ through either altering the gate time, or an AC Stark shift can be applied to introduce a compensating detuning.

% Beyond $N=2$, $k$ atoms in $|r_2\rangle$ leads to the efficient generation of the family of Dicke states, $|D^N_k\rangle$, which would otherwise be prepared in $\mathcal{O}(k\;\text{log}(n/k))$ \cite{dicke_state_prep}.

\subsection{\label{sec:cz}Gates using a driven cavity}
\begin{figure}
    \centering
    \includegraphics[width=8.4cm]
    {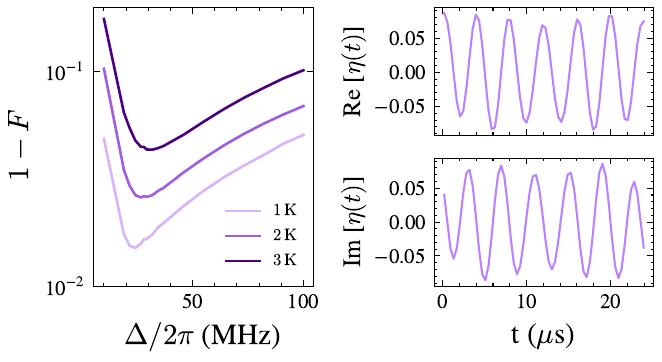}
    \caption{ The left panel shows the gate infidelity as a function of detuning for three candidate temperatures. Increasing temperature broadens the distribution of photon-number-dependent phases and reduces the fidelity as per Eq.~(12). The infidelity initially decreases with detuning as the residual dispersive dressing is suppressed, before the decay events become comparable. The right panels show the real and imaginary quadratures of the GRAPE-optimised cavity drive for $\Delta= 2\pi \times25\mathrm{MHz}$ and $T=1\mathrm{K}$, in units of MHz.}
    \label{fig:thermalphotons}
\end{figure}

We now adapt the protocol found in Ref.~\cite{Jandura2024} to our four level system and a thermal cavity and find optimal cavity drives, as seen in Fig. (\ref{fig:schematics})(b). The protocol is as follows, population is transferred to $|r_2\rangle$ by a $\pi$ pulse on the
$|1\rangle\leftrightarrow\!|r_2\rangle$ transition under
\begin{equation}
  H_{\mathrm{excite}}(t)=\sum_{j}\frac{1}{2}\!\left(
      \Omega_j^{1,r_2}(t)\,e^{i\phi_2}\,|r_2\rangle\langle 1|_j
      +\mathrm{H.c.}\right).
  \label{eq:Hexcite_cz}
\end{equation}
The cavity is then driven by a classical microwave field,
\begin{equation}
  H_{\mathrm{drive}}(t)
  = i\!\left[\eta(t)e^{-i\omega_d t}-\eta^{*}(t)e^{i\omega_d t}\right]
    (\hat a^{\dagger}+\hat a).
  \label{eq:Hdrive_cz}
\end{equation}
with complex envelope $\eta(t)$, the lab frame Hamiltonian is then simply $H_{TC}  + H_{drive}$. Moving to the frame rotating at $\omega_d$, using the transformation
$\hat U_r=\exp[i\omega_d t(\hat a^{\dagger}\hat a+\sum_j|r_2\rangle\langle r_2|_j)]$ defines the detunings
\begin{equation}
  \delta=\omega_c-\omega_d,\qquad
  \mu=\omega_q-\omega_d.
  \label{eq:detunings_cz}
\end{equation}
where $\delta$ is a detuning between drive and cavity and we also define the parameter $r= g/|\Delta|$, a full derivation can be found in Appendix B. Displacing the cavity to absorb the drive, $D(\alpha) = \mathrm{exp}(\alpha\hat{a}^\dagger-\alpha^* \hat{a})$, and adiabatically eliminating
$|r_1\rangle$ gives
\begin{equation}
  \begin{split}
  H_{\mathrm{eff}}=\;&\delta\,\hat a^{\dagger}\hat a+\big(\zeta(t)\hat a^{\dagger}+\zeta^{*}(t)\hat a\big)\hat n\\
     &+\chi\,\hat n\,\hat a^{\dagger}\hat a
     +\chi\big(1+|\alpha(t)|^{2}\big)\hat n.
  \end{split}
  \label{eq:Heff_cz}
\end{equation}
where $\hat n=\sum_j|r_2\rangle\langle r_2|_j$ the Rydberg population operator, $\alpha(t)$ is the classical cavity amplitude and $\zeta(t)=\chi\,\alpha(t)$ is a state-dependent force generated through displacing the cross-Kerr coupling. The second term drives the gate, each $\hat n$ sees a linearly driven oscillator, whose evolution generates a geometric phase $\exp(-i\Theta\hat n^{2})$ with $\Theta$ the enclosed phase-space area. The third is a residual cross-Kerr term $\chi\hat n\hat a^{\dagger}\hat a$, and is non negligible in a thermal cavity as it gives a phase proportional to the photon number. This gives two conditions, firstly the cavity must
disentangle from the qubits which requires the drive spectrum to vanish at each frequency $\omega_n=\delta+n\chi$. Secondly the residual term generates $\exp(-i\hat n\hat a^{\dagger}\hat aX)$ with $X=\int_0^{t_g}\chi\,dt$ and imprints the phase $-nkX$ on a Fock state $|k\rangle$, and must vanish for every $k$ imposing the gate time, $t_g=2\pi/{\chi}$. An initial coherent amplitude evolves as $\alpha(0)\;\mathrm{exp}[{-i(\delta+n\chi)t_g}]$ and returns to itself in every sector only if $\chi t_g=2\pi$, loop closure for an arbitrary initial cavity state and reduction of residual Kerr effects both occur at $t_g$. When both conditions hold the two-qubit evolution in the $\{|0\rangle, |r_2 \rangle \} $ basis is therefore
\begin{equation}
  U=\mathrm{diag}\!\left(1,\;e^{-i\Theta},\;e^{-i\Theta},\;e^{-i4\Theta}\right).
  \label{eq:Ugeom}
\end{equation}
The corrective single-qubit rotations can then be absorbed into the excitation and de-excitation pulses in Eq.~\eqref{eq:Hexcite_cz}, the phase may be chosen such that $\phi=\Theta$, giving the ideal gate $U_{\mathrm{CZ}}=\operatorname{diag}(1,1,1,-1))$. Choosing $t_g = 2\pi/\chi$ ensures the the phase gained in on loop is an integer number of $2\pi$ for every photon number. The exact dispersive shift is however nonlinear, with the first correction giving $-\chi^2/{\Delta}(k+1)^2$. This means the gate time cannot cancel both terms, and different photon numbers acquire different phases. A separate error channel arises from the residual population of $| r_1\rangle$, which is of order $r^2(\bar n+1)$, and arises from the strong coupling between the transition and the cavity. Collecting both contributions the infidelity obeys the scaling
\begin{equation}
  1 - F \;\propto\; r^2(\bar n + 1) \;+\; r^4\,\mathrm{Var}\!\left[(k+1)^2\right].
  \label{eq:floor}
\end{equation}
Because only $|r_2\rangle$ is genuinely populated, Rydberg decay enters primarily through
$\Gamma_{r_2}$, but the gate is subject to all other decays previously discussed. Combining the gate time and \eqref{eq:floor}, the coupling required to reach a target
infidelity $\varepsilon$ within a gate time $t_g$ is
\begin{equation}
  \frac{g}{2\pi}\;\gtrsim\;\frac{1}{t_g}\sqrt{\frac{\bar n+1}{\varepsilon}}.
  \label{eq:designrule}
\end{equation}
We simulate the full driven Tavis-Cummings Hamiltonian with loss and find the optimal pulse shape, $\eta(t)$, using the GRAPE optimal control method \cite{Khaneja2005}. The fidelity as a function of the detuning at different temperatures is shown in  Fig.~(\ref{fig:thermalphotons}). The infidelity initially falls as $r^2$ with increasing $\Delta$, before the decay accumulated over the longer gate duration becomes comparable. At fixed $\Delta$, increasing temperature broadens the distribution of $k$ dependant phases and hence reduces the fidelity.

\section{\label{raman}Bichromatic Raman gate}
\subsection{Two qubit gates}
\begin{figure}
    \centering
    \includegraphics[width=8.4cm]{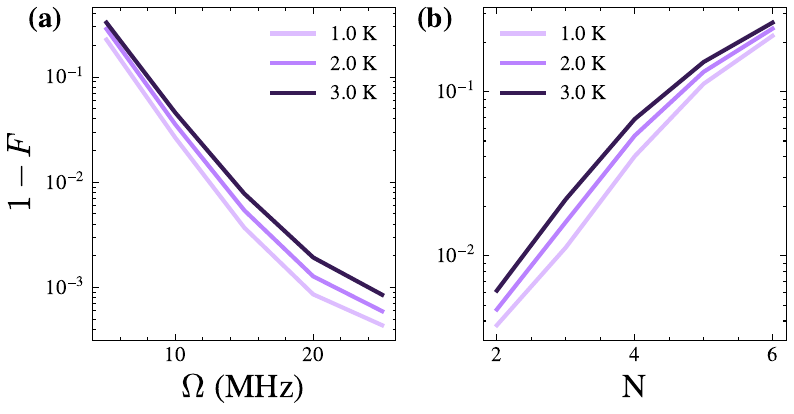}
    \caption{(a) Gate infidelity against drive strength, $\Omega$, for temperatures T = $1,2,3$K (b) GHZ state infidelity for N atoms, with more atoms the probability of a decay event or phase error scales with N}
    \label{fig:brgate}
\end{figure}

Both preceding protocols require real population of the Rydberg manifold and hence suffer from the complete Rydberg decay rate. We now show that a controlled-phase gate can instead be generated by two detuned optical Raman drives while the Rydberg states remain only virtually populated. As in the driven-cavity protocol, only the $|1\rangle$ state is excited. It is coupled off-resonantly to both $|r_1\rangle$ and $|r_2\rangle$, with detunings $ \Delta_1=\omega_{r_1}-\omega_{L_1},\; \Delta_2=\omega_{r_2}-\omega_{L_2},
$ where $\omega_{L_1}$ and $\omega_{L_2}$ are the respective drive frequencies. A schematic of the gate is shown in Fig.~(\ref{fig:schematics})(c). The drive term is then 
\begin{equation}
\begin{split}
    H_{\mathrm{Raman}}(t) = \sum_{j} \frac{1}{2}\!(
        \Omega_j^{1,r_1}(t)\,e^{i(\phi_1-\omega_{L_1}t)}|r_1\rangle\langle 1|\\
        +\Omega_j^{1,r_2}(t)\,e^{i(\phi_2-\omega_{L_2}t)}|r_2\rangle\langle 1|
        + \mathrm{H.c} ).
\end{split}
\label{eq:Hraman}
\end{equation}
The complete laboratory-frame Hamiltonian is therefore
\(H_{\mathrm{TC}}+H_{\mathrm{Raman}}(t)\). We move to the rotating frame of the drives, defined as
\begin{equation}
\begin{split}
  \hat U_r = \exp\Big\{ i t \Big[
    \omega_{L1}\textstyle\sum_j \hat n_{r_1}
  + \omega_{L2}\sum_j \hat n_{r_2}\\
  + (\omega_{L2}-\omega_{L1})\, \hat a^\dagger \hat a \Big]\Big\}.
  \label{eq:crf-frame}
\end{split}
\end{equation}
we now define $\delta_r = \omega_c - (\omega_{L2}-\omega_{L1})$, analogous to Eq. \eqref{eq:detunings_cz}, and impose the constraint $\Delta_2 - \Delta_1 = \Delta+\delta_r$ to ensure frame consistency. In this frame, the Hamiltonian becomes
\begin{equation}
\begin{split}
  H =  \delta_r \hat a^\dagger \hat a
  + \sum_j \Big[ \Delta_1 \hat n_{r_1} + \Delta_2 \hat n_{r_2}
  + g\big( \sigma^{+}_{j}\hat a + {\rm H.c.} \big) \\
  + \tfrac{1}{2}\big(\Omega_1 \big( \sigma^{r_1,1}_{j} + {\rm H.c.}\big)
   + \Omega_2 \big( e^{-i\phi}\sigma^{r_2,1}_{j} + {\rm H.c.}\big)\big)
  \Big].
\end{split}
  \label{eq:crf-frameham}
\end{equation}
where $\phi = \phi_2 - \phi_1$ is the relative optical phase and we define $\sigma_j^{1,r_1} = |1\rangle \langle r_1|$. Only the difference of the two laser phases appears, so the drives must be
phase-locked. We consider two regimes, a strong drive regime where $\Omega_i \gg \Delta_i$, and a weak drive regime where $\Omega_i/\Delta_i \ll 1$. In the weak drive regime we can eliminate the Rydberg manifold through a Schrieffer-Wolff transformation (Appendix C), and get the effective Hamiltonian
\begin{equation}
  H_{eff} = \delta_r \hat{a}^\dagger \hat{a}+ \hat n_1
  \left(\lambda(t) e^{i\phi}\hat a +\lambda^* (t) e^{-i\phi}\hat a^\dagger \right).
  \label{eq:crf-force}
\end{equation}
where $\lambda(t) = g\,\Omega_1 \Omega^*_2/4\Delta_1 \Delta_2$ and $\hat n_1 = \sum_j | 1\rangle\langle 1|_j$ for the weak drive regime and $\lambda(t) \rightarrow g/2$ for a strong drive. The force is dependant on the number of atoms in the logical state $|1\rangle$, analogous to the force term $\zeta(t)\hat a^\dagger \hat n$ of Eq. \eqref{eq:Heff_cz}. Introducing the displacement $\alpha_n=-\lambda^* n/\delta_r$ gives
\begin{equation}
D^\dagger(\alpha_n)H_nD(\alpha_n)=\delta_r a^\dagger a-\frac{|\lambda|^2n^2}{\delta_r}.
\end{equation}
Through propagating $H_{eff}$, the first two Magnus terms describe the conditional displacement $\beta_n(t)=\lambda n/{\delta_r}\left(1-e^{-i\delta_r t}\right)$ 
and a geometric phase $\Theta_n(t)=\lambda^2n^2/{\delta_r^2}\left(\delta_r t-\sin\delta_r t\right)$. In the weak drive regime one loop is completed at $t_g=2\pi/|\delta_r|$, and for the strong drive regime it approaches $t_g \rightarrow2\pi/g$, giving the evolution
\begin{equation}
U(t_g)=\exp\!\left[2\pi i\left|\frac{\lambda}{\delta_r}\right|^2\hat n_1^2\right].
\end{equation}
The error resulting from Rydberg decay is the product of the decay rate, Rydberg population and gate time, which simplifies to 
% For equal Rydberg decay rates $\Gamma_r$, the error is therefore a function of the Rydberg admixture ($p_R = |\Omega_1|^2/{4\Delta_1^2}
%     +|\Omega_2|^2/{4\Delta_2^2}$), and the integrated Rydberg population is $2\pi/g$ giving
\begin{equation}
\varepsilon_{\mathrm{Ryd}}\simeq \frac{2\pi\Gamma_r}{g}.
\end{equation}
Increasing the drive strength reduces the gate duration but increases the instantaneous Rydberg admixture by the same factor. At the CZ operating point $|\lambda/\delta_r|^2 = 1/4$, the sectors
containing $n$ and $m$ atoms in $|1\rangle$ follow trajectories separated by
\begin{equation}
    |\beta^{(S)}_n(t)-\beta^{(S)}_m(t)|^{2}
    = \frac{(n-m)^{2}}{4}\,|1-e^{-i\delta_r t}|^{2}.
\end{equation}
We then use $\int_0^{t_g}|1-e^{-i\delta_rt}|^2dt=2t_g$ and averaging the resulting $e^{i\kappa(n-m)^2}$ over the the two qubits gives the thermal dephasing scaling $\varepsilon_{\mathrm{cav}}\simeq\kappa(2\bar n+1)t_g/5$. At the balanced point ($\Omega_1/\Delta_1 = \Omega_2/\Delta_2$), the leading cross-Kerr term is vanishes and the remaining photon-number dependence enters only through higher-order corrections to the force amplitude. Since infidelity is quadratic in phase error, this contributes $\varepsilon_{\mathrm{th}}=O\!\left[r^4\bar n(\bar n+1)\right]$, the full error scaling is therefore
\begin{equation}
    1-F\propto\frac{2\pi\Gamma_r}{g}+\frac{1}{5}\kappa(2\bar n+1)t_g+O\!\left[r^4\bar n(\bar n+1)\right].
\end{equation}
The gate condition $tg = 2\pi/\delta_r$ is under the assumption of a constant force, however loop closure can be achieved through pulse shaping alone, so in principle any arbitrary gate time can be chosen, subject to pulse shape constraints. The simulated fidelities as a function of drive strength can be found in Fig. (\ref{fig:brgate}), for the strong drive regime with $\Delta_1 = 18.89$MHz and $\Delta_2 = -12.82$MHz we find a lossy fidelity of $F = 0.997$.
\subsection{N qubit gates}
This gate can be naturally extended to N qubits, to create useful resources such as GHZ states. To prepare an $N$-qubit GHZ state, all atoms are initialized in
$\ket{+}^{\otimes N}$ and subjected to the same drive. The atoms therefore undergo $U_{\mathrm{GHZ}}=\exp\left(\frac{i\pi}{2}\hat n_{1}^{2}\right)$, expanding $n_{1}^2 $ gives
\begin{equation}
    U_{\mathrm{GHZ}}
    =
    \left(\prod_{j=1}^{N}S_j\right)
    \left(\prod_{j<k}\mathrm{CZ}_{jk}\right).
\end{equation}
Where $S$ can be corrected by single qubit rotations and the second term applies CZ gates between qubits $j$ and $k$ making a graph state. In particular,
\begin{equation}
    U_{\mathrm{GHZ}}\ket{+}^{\otimes N}
    =
    \frac{
        e^{i\pi/4}\ket{+}^{\otimes N}
        +
        e^{-i\pi/4}\ket{-}^{\otimes N}
    }{\sqrt{2}},
\end{equation}
which is an $X$-basis GHZ state. Applying Hadamard gates to all atoms and a known single-qubit phase correction produces the conventional GHZ
state. As the interaction is permutation symmetric, all states containing $k$ atoms in $|1\rangle$ follow the same evolution. We group the basis states according to this Hamming weight and define $M_{kl}$ as the complex density-matrix element between any basis states of weights $k$ and $l$.
\begin{equation}
    F_{\mathrm{GHZ}}
    =
    \max_{\beta}
    \frac{1}{4^{N}}
    \sum_{k,l=0}^{N}
    \binom{N}{k}\binom{N}{l}
    e^{-i\varphi_k(\beta)}
    M_{kl}
    e^{i\varphi_l(\beta)},
\end{equation}
where $\varphi_k(\beta)= \frac{\pi}{2}k^{2}+\beta k$ and $\binom{N}{k}\binom{N}{l}$ counts the density matrix elements connecting states of weight $k$ and $l$. The maximisation over $\beta$ removes a correctable uniform single-qubit $Z$ phase. Imperfect loop closure suppresses the coherence
between sectors $k$ and $l$ by
$\exp[-(2\bar n+1)|\alpha(t_{\mathrm g})|^{2}(k-l)^{2}/2]$.
Cavity loss produces an additional correlated dephasing contribution proportional to
$\kappa(2\bar n+1)(k-l)^{2}
\int_{0}^{t_{\mathrm g}}|\alpha(t)|^{2}\,dt$, while Rydberg decay causes
leakage determined by the time integrated Rydberg population. The relationship between N and $1-F$ can be seen in Fig. (\ref{fig:brgate})(b).
\subsection{Parallel gates}
It is also desirable to perform multiple gates in parallel, this can be achieved through choosing different $\delta_r$ for different pairs. We denote the two pairs as $p=AB$ and $p=CD$ and define the operators
\begin{equation}
    \hat n_{AB}=\hat n_{1,A}+\hat n_{1,B},
    \qquad
    \hat n_{CD}=\hat n_{1,C}+\hat n_{1,D}.
    \label{eq:parallel-population-operators}
\end{equation}
We can eliminate the Rydberg manifold and get the effective Hamiltonian
\begin{equation}
\begin{split}
    H_{\mathrm{eff}}^{\mathrm{pair}}(t)
    ={}& \sum_p
    \hat n_p
    \left[
        \xi_p(t)\hat a+\xi_p^*(t)\hat a^\dagger
    \right]
    \\
    &+\sum_p\epsilon_p(t)\hat n_p,
    \label{eq:parallel-effective-hamiltonian}
\end{split}
\end{equation}
where $\xi_p(t)= e^{i\phi_p}\lambda_p(t)e^{-i\delta_p t}$ and $\epsilon_p(t)=-|\Omega_{1,p}(t)|^2/{4\Delta_{1,p}}-|\Omega_{2,p}(t)|^2/{4\Delta_{2,p}}$. The pairwise displacements and self-geometric phases are
\begin{equation}
    \beta_p(t)
    =-i\int_0^t \xi_p^*(s)\,ds,
    \label{eq:parallel-displacements}
\end{equation}
\begin{equation}
    \Theta_p(t)
    =\operatorname{Im}
    \int_0^t ds\int_0^s ds'
    \xi_p^*(s)\xi_p(s'),
    \label{eq:parallel-self-phases}
\end{equation}
and the unwanted inter pair phase is
\begin{multline}
    \Theta_X(t)
    =\operatorname{Im}
    \int_0^t ds\int_0^s ds'
    \bigl[
        \xi_{AB}^*(s)\xi_{CD}(s')
        \\
        +\xi_{CD}^*(s)\xi_{AB}(s')
    \bigr].
    \label{eq:parallel-cross-phase}
\end{multline}
The cavity disentangles from all four atoms when
\begin{equation}
    \beta_{AB}(t_g)=\beta_{CD}(t_g)=0.
    \label{eq:parallel-closure}
\end{equation}
\begin{figure}
    \centering
    \includegraphics[width=8.4cm]{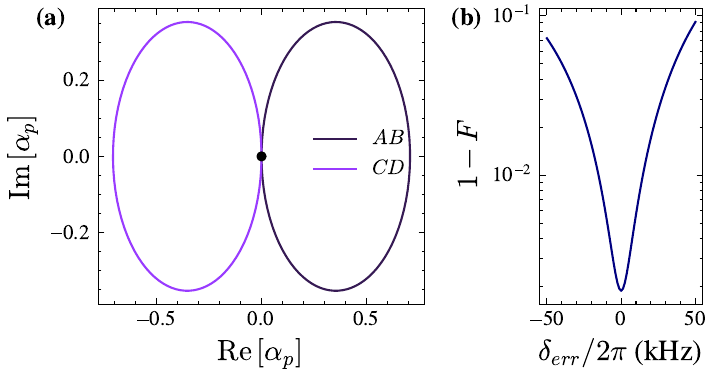}
    \caption{(a) Cavity displacements generated by the bichromatic Raman gate. Counter-rotating loops occur when choosing opposite $\delta_r^p$ configuration for the AB and CD pairs allowing for parallel operation in the same mode (b) Average gate fidelity as a function of an absolute error applied to $\delta_r^{CD}$, leading to an asymmetry.}
    \label{fig:placeholder}
\end{figure}
We therefore require $\Theta_{AB}(t_g) = \pi/2,\;\Theta_{CD}(t_g) = -\pi/2$ and $\Theta_X = 0$. Loop closure occurs for each pair at $t_g = 2\pi m_n/\delta_r^p$, where $m_n$ is any integer. $m_n$ can be chosen to be of opposite sign for the different pairs, ie $m_n^{AB} = 2,\;m_n^{CD} = -2$, such that they complete loops in counterpropogating directions and suppresses cross geometric phase. Taking $t_g = 2.5 \mu s$ and the $m_n$ values above we get $\delta_r^{AB} = 2\pi \times 0.8MHz$ and $\delta_r^{CD} = -2\pi \times 0.8MHz$. In the full four-level model, finite-detuning produces a residual mixed-sector phase. This term is removed by rotating the
relative optical phase of the $CD$ Raman pair. If $\xi_{CD}(t)\rightarrow e^{i\varphi}\xi_{CD}(t)$, the individual
displacements and self-phases are unchanged up to a rigid rotation, whereas
$\Theta_\times$ changes with $\varphi$. For the independently optimized pulses used in Fig. (\ref{fig:placeholder}), the required correction is $\varphi=\pi/2$. We simulate the full open system and find a fidelity of $F = 0.994$. 

\section{\label{sec:toric}non local error correction}
A central motivation for non-local entangling gates is their
potential to enhance connectivity and enable long-range interactions required in quantum
error-correcting codes with non-local stabilisers, and more specifically
by any code whose syndrome-extraction circuit requires interactions
beyond nearest-neighbours. Cavity-assisted syndrome extraction
has recently been analysed at the circuit level in
Ref.~\cite{Chandra2025}, which showed that cavity mediated gates
support fault-tolerant thresholds even when the cavity error exceeds
the local gate error by an order of magnitude. In this section we consider the wall clock time of error correction rounds. We use the constraint that cavity operations must be applied sequentially, so we must consider both the reach of the gate and the number of non-local gates a code demands per round. Firstly, we consider the toric code, where the non-local stabilisers arise only at the periodic boundaries, and propose a
hybrid protocol where local Rydberg gates are used for the bulk syndrome extraction and cavity mediated gates are used for the periodic boundaries. We then extend this to the wider class of bivariate bicycle codes, which have higher encoding rates and more complex long-range connectivity requirements.
\subsection{Toric code}
The toric code is defined on a $2L^2$-qubit register, with data qubits arranged on the links of an $L \times L$ square lattice subject to periodic boundary conditions in both directions, giving $[[2L^2,2,L]]$. Each plaquette of the lattice hosts a weight-four $Z$-type stabiliser generator $A_p = \prod_{i \in p} Z_i$, and each vertex hosts a weight-four $X$-type stabiliser generator $B_v = \prod_{i \in v} X_i$, with the two families of stabilisers measured via ancilla qubits placed at the corresponding plaquette/vertex. For a plaquette in the bulk, the four data qubits in the support of the stabiliser generator are located at nearest-neighbour sites on the array, and the stabiliser generator is measured by a standard sequence of four CZ gates between the ancilla and each of its four neighbours, followed by ancilla readout. This sequence is local in the sense that every entangling gate acts between physically adjacent atoms, and can therefore be implemented directly using the Rydberg blockade with minimal transport. The periodic boundary conditions, however, identified at the edges of the lattice, have the stabilisers straddling the boundary acting on data qubits separated by a distance $\sim L\,d$, where $d$ is the inter-atom spacing within the array. Satisfying these boundary stabilisers requires either physically transporting the ancilla atoms across the full extent of the array, or abandoning the periodic boundary condition in favour of an open or fixed boundary. A single error-correction round therefore follows two sequential stages. In the first stage, every ancilla associated with a bulk stabiliser sequentially performs CZ gates with its four nearest-neighbour data qubits which may, in principle, be parallelised across non-overlapping plaquettes/vertices, with time to complete this stage $T_{\rm int}$. In the second stage, the periodic boundary stabilisers are closed by entangling each boundary data qubits with the ancillas, denoted as $T_{\rm PBC}$, which we evaluate separately for the shuttling and cavity-mediated implementations below. The total duration of one error-correction round is then $T_{\rm round} = T_{\rm int} + T_{\rm PBC}$. Since $T_{\rm int}$ is identical for both implementations, interior syndrome extraction is unaffected by how the boundary is closed, any speed-up obtained from the cavity-mediated gate is entirely through a reduction of $T_{\rm PBC}$. We use the array reconfiguration formula in Ref.~\cite{Xu2024}, in which $T_{\rm PBC}$ can be defined in terms of the Manhattan distance. For the toric code this would be
\begin{equation}
    T_{\rm PBC}^{\rm shuttle} = 2\left(\sqrt{6}+\sqrt{3}\right)\sqrt{\frac{2dL}{a}},
    \label{eq:Tpbc_shuttle}
\end{equation}
As cavity-mediated gates must be applied sequentially, the total number of sequential cavity gate applications scales as $4L$. The resulting boundary-closure time is therefore
\begin{equation}
    T_{\rm PBC}^{\rm cavity} = 4 L\, T_g,
    \label{eq:Tpbc_cavity}
\end{equation}
where $T_g$ is the CZ gate time. In contrast to the shuttling protocol, Eq.~\eqref{eq:Tpbc_cavity} scales linearly, $\mathcal{O}(L)$, rather than as $\mathcal{O}(\sqrt{L})$, the cavity-mediated approach is therefore not asymptotically advantageous for arbitrarily large $L$ it can however gain a practical speedup for typical array sizes ($L<100$). We use the gate times for the bichromatic Raman gate which can be chosen arbitrarily by the pulse shape $T_g = \{1.5,2.5,3.5\} \mu s$. Comparing Eqs.~\eqref{eq:Tpbc_shuttle} and~\eqref{eq:Tpbc_cavity}, the cavity-mediated protocol yields a shorter boundary-closure time, $T_{\rm PBC}^{\rm cavity} < T_{\rm PBC}^{\rm shuttle}$, whenever the cavity gate time satisfies
\begin{equation}
    T_g(L) < \frac{\sqrt{6}+\sqrt{3}}{2}\sqrt{\frac{2d}{La}}.
    \label{eq:breakeven}
\end{equation}
Fig.~(\ref{fig:cavqec}) shows $T_{\rm round}$ as a function of array length $L$ for the all-shuttling protocol and for the hybrid protocol at several representative gate times. For an array of size $L = 50$, and taking $T_g = 2.5~\mu\mathrm{s}$, the boundary-closure stage alone is accelerated from $T_{\rm PBC}^{\rm shuttle} = 1.3~\mathrm{ms}$ to $T_{\rm PBC}^{\rm cavity} = 0.5~\mathrm{ms}$. Once considering the interior syndrome-extraction time, the hybrid protocol accelerates the full error-correction round by $2.3\times$ relative to the all-shuttling protocol. The speed-up is largest in the intermediate-$L$ regime where the $\mathcal{O}(\sqrt L)$ shuttling overhead has become a non-negligible fraction of the round time but is not yet so large as to dominate it entirely.
\subsection{Bivariate bicycle codes}
\begin{figure}
    \centering
    \includegraphics[width=8cm]{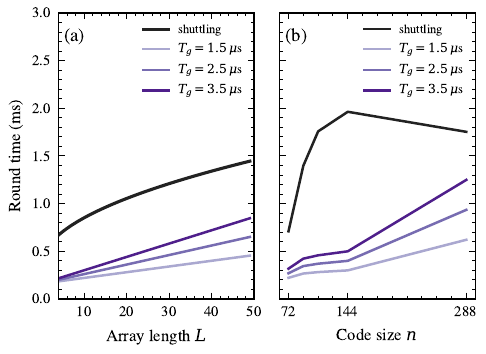}
    \caption{Time to complete one error-correction round for the
    all-shuttling protocol (black) and the hybrid cavity protocol at different $T_g$. Atom rearrangement
    parameters follow Ref.~\cite{Xu2024}, with grid spacing $d=5~\mu$m
    and peak acceleration $a_p=0.02~\mu\mathrm{m}/\mu\mathrm{s}^2$.
    (a)~Toric code versus array length $L$, the boundary stabilisers are
    closed by transport, or sequentially by the cavity, at $L=50$ and $T_g=2.5~\mu$s the hybrid protocol shortens
    the round by a factor of 2.3. (b) Bivariate-bicycle codes of block length $n$, cavity rounds last $T_{\mathrm{int}}+n_{\mathrm{ops}}T_g$ with ($n_{\mathrm{ops}}=48$, $78$, $90$, $102$, $342$), while
    the shuttling baseline closes the same wrapped edges with one parallel movement for each length.}
    \label{fig:cavqec}
\end{figure}
The toric code is the simplest member of a wider class of codes whose
stabilisers are non-local in any planar layout. Quantum low-density
parity-check codes, in particular the bivariate
bicycle (BB) \cite{Bravyi2024} and lifted-product codes
\cite{Panteleev2021}, have high encoding rates at the
price of non-local stabilisers. A BB code on a $\ell\times m$ grid is specified by two three-term
polynomials $A(x,y)$ and $B(x,y)$ in commuting cyclic shifts, with parity check matrices
$H_X=[A\,|\,B]$ and $H_Z=[B^{\mathsf T}|\,A^{\mathsf T}]$. In the layout of Ref.~\cite{Bravyi2024} each cell hosts one data qubit of each sublattice and one ancilla of each
type, and every Tanner edge is a displacement $(a,b)$ given by a
monomial, of at most $s_{\max}$ cells. The non-locality is therefore entirely a boundary effect of flattening the torus onto a physical array, an edge of class $(a,b)$ wraps for exactly $|a|m+|b|\ell-|a||b|$ of its $\ell m$ instances, a wrapped
edge spans $\ell-|a|$ (or $m-|b|$) cells, and only edges whose planar
extent exceeds the interior range $R=s_{\max}$ already required by
the code's unwrapped edges must be closed non-locally. Summing over
the edge classes of both stabiliser families bounds the per-round bus
load by a perimeter law,
\begin{equation}
  n_{\mathrm{bus}} \;\le\; 2\!\!\sum_{p\,\in A\cup B}\!\!
  \bigl(|a_p|\,m+|b_p|\,\ell)
  \label{eq:nbus}
\end{equation}
with the toric code recovered as the two-monomial case $A=1+x$,
$B=1+y$, for which the count returns $n_{\mathrm{ops}}=4L$. For the code family of Ref.~\cite{Bravyi2024}, the exact loads are $n_{\mathrm{ops}}=48$, $78$, $90$,
$102$ and $342$ for the $n=72,90,108,144,288$ codes respectively. Every non-local edge is then closed by a cavity mediated CZ applied sequentially $T_{\mathrm{nl}}^{\mathrm{cavity}} = n_{\mathrm{ops}}\,T_g,$. Fig. (\ref{fig:cavqec}) compares the resulting round time for different size codes, the hybrid protocol shortens the round
by a factor $2.6$--$4.8$ across the four smaller codes, narrowing to $1.7$ for $[[288,12,18]]$ as there are fewer rounds of shuttling and fewer stabiliser with components that cross the boundary. Unlike the toric code, the cavity mediated gates are not a vanishing fraction of
a round, they provide $\phi=11$--$20\%$ of all two-qubit gates across the family, so the mean two-qubit error is
$\bar p_2=\varepsilon_{\mathrm{loc}}
+\phi\,(\varepsilon_{\mathrm{cav}}-\varepsilon_{\mathrm{loc}})
\approx1.3$--$1.6\times10^{-3}$ for
$\varepsilon_{\mathrm{cav}}=4\times10^{-3}$ and
$\varepsilon_{\mathrm{loc}}=10^{-3}$, lies below the threshold of $\approx0.7\%$ reported for the BB family under the stochastic noise model of Ref \cite{Bravyi2024}. Codes with higher weight non-local stabilizer generators, would benefit further from multi-qubit gates, whose cost per check is independent of both its weight and its reach. \\

The cavity introduces temporally correlated noise that is not captured by an independent Pauli channel. If the photon number $k$ remains fixed over a sequence of gates, the interaction-picture error accumulated during a round is, to first order
\begin{equation}
    U_{\mathrm{err}}^{(\mathrm{round})}\simeq
\exp\!\left[
-\frac{i}{2}
\sum_{j=1}^{n_{\mathrm{ops}}}
\epsilon_j(k)\tilde P_j
\right]
\end{equation}
where $\epsilon_j(k)$ is the photon-number-dependent phase error of gate $j$, and $\tilde P_j$ is its error generator propagated through the ideal circuit. The resulting infidelity scales as $n_{\mathrm{ops}}^{2}\epsilon_k^{2}$ only in the worst-case limit in which identical errors add coherently. Circuit propagation, non-commuting operations, compilation, and stabiliser measurements can instead suppress or restructure this accumulation. Thermal excitation and photon loss introduce a distinct error by changing $k$ during the round. A jump between gates changes the phase calibration of subsequent operations, whereas a jump during a gate can additionally prevent closure of the cavity trajectory and leave residual data cavity entanglement. Tracing over the unobserved cavity then produces temporally correlated dephasing.

\section{\label{sec:conclusion}Outlook}
In this work we have outlined multiple cavity mediated gates for neutral atoms coupled to a three dimensional microwave cavity. We firstly adapted established protocols to a 4 level architecture and derived bounds on fidelity for a thermal cavity. We then introduced a bichromatic Raman gate which generates entanglement between two atoms while only virtually populating the Rydberg states, achieving fidelities of 0.997. Finally we showed the application of these gates in error correcting codes, and that through using cavity mediated gates, the wall clock time of the gross code can be sped up by a factor of 4.8. While the bichromatic Raman gates can in principle be parallelised, it comes at a cost of lower fidelity. Other routes to greater parallelisability exist, the cavity could be engineered to host multiple modes, then extra Rydberg states $\{|r_3\rangle,|r_4\rangle \}$ could be used. Overlapping cavity arrays have also been proposed as a route towards having individually addressable sites rather than only the global control currently offered. Alternatively the global control can also be used as a powerful resource to engineer large cluster states for measurement based quantum computing, which can be combined with parallel, local Rydberg gates for efficient construction. Using bosonic modes as an entanglement bus is a promising direction for quantum computing architectures, and provided greater parallelisability can bypass the connectivity challenges associated with larger arrays. The mode itself can also be used as a resource, future work could apply developments in hybrid oscillator qubit models to this approach or the potential applications in analog quantum simulation.

\section*{Acknowledgments}
M.J.H.K, M.B.S and C.J.W are supported by the Engineering and Physical Sciences Research Council (EPSRC) (Grant No EP/Y035046/1).

% \section*{Author contributions}

% \section*{Data availability}
% The data that support the findings of this study are available from the corresponding author upon reasonable request.
\bibliography{bibliography}
\setcounter{subsection}{0}
\section*{Appendix A: Dispersive Tavis cummings}
We write the Hamiltonian as the sum of the bare Hamiltonian of the cavity and qubits and the interaction $H = H_0 + H_i$, where $H_0 = \omega_c a^\dagger a  + \tfrac{1}{2} \sum_j^N \omega_j \sigma^z_j$ and $H_i = \sum_j^N g_j(\sigma^+_j a + \sigma^-_j a^\dagger)$.
We now apply a Schrieffer-Wolff-Luttinger-Kohn unitary transformation $U = e^S$ to eliminate the first-order interaction terms. The effective Hamiltonian is expanded using the Baker-Campbell-Hausdorff relation: 
\begin{equation}
\begin{split}
    \tilde{H} = e^S H e^{-S} = H + [S,H] + \\\frac{1}{2}[S,[S,H]] + \frac{1}{3!}[S,[S,[S,H]]] + \dots
\end{split}
\end{equation}
To cancel the interaction term $H_i = \sum g_j (\sigma^+_j a + \sigma^-_j a^\dagger)$, we choose an anti-Hermitian generator $S$ such that $[S, H_0] = -H_i$. The generator $S$ that satisfies this is:
\begin{equation}
    S = \sum_j^N \frac{g_j}{\Delta_j} (\sigma^+_j a - \sigma^-_j a^\dagger)
\end{equation}
Substituting $S$ back into the expansion, the first-order terms cancel out ($H_i + [S, H_0] = 0$), leaving the second-order term, $\tilde{H} \approx H_0 + \tfrac{1}{2}[S, H_i]$.
Calculating the commutator:
\begin{align}
\frac{1}{2}[S, H_i] &= \frac{1}{2} \left[ \sum_j^N \frac{g_j}{\Delta_j} (\sigma^+_j a - \sigma^-_j a^\dagger), \sum_j^N g_j(\sigma^+_j a + \sigma^-_j a^\dagger) \right]     
\end{align}
 leads to the transformed Hamiltonian in the dispersive regime
\begin{equation}
\begin{split}
    \tilde{H} = (\omega_c + \sum_j^N \chi_j \sigma^z_j) a^\dagger a + \tfrac{1}{2} \sum_j^N (\omega_j + \chi_j) \sigma^z_j + \\ \sum_{j< k} \frac{g_jg_k}{2} \left( \frac{1}{\Delta_j} + \frac{1}{\Delta_k} \right)  (\sigma^+_j \sigma^-_k + \sigma^-_j \sigma^+_k)
\end{split}
\end{equation}
where $\chi_j = g_j^2/\Delta_j$.
Here we can see that the cavity frequency is shifted to 
$$
\tilde{\omega}_c = \omega_c + \sum_j^N \chi_j \sigma^z_j,
$$
depending on the inversion of each qubit, and each qubit frequency is shifted
$$
\tilde{\omega}_j = \omega_j + \chi_j
$$
and there is a term
$$
J_{jk} = \frac{g_j g_k}{2} \left( \frac{1}{\Delta_j} + \frac{1}{\Delta_k} \right)
$$
that entangles the qubits. 

\section*{\label{app:cz}Appendix b: driven cavity gates}
 
\subsection{Displaced frame and adiabatic elimination}
 
The drive of Eq.~\eqref{eq:Hdrive_cz} is removed by displacing the cavity,
$\tilde\rho=D(\alpha)^{\dagger}\rho\,D(\alpha)$ with
$D(\alpha)=\exp(\alpha\hat a^{\dagger}-\alpha^{*}\hat a)$, provided the classical
amplitude obeys the driven-cavity equation of motion, including photon loss at
rate $\kappa$,
\begin{equation}
  \dot\alpha=-\big(i\delta+\kappa/2\big)\alpha+\eta(t),
  \qquad \alpha(0)=0 .
  \label{eq:alpha}
\end{equation}
This cancels every term linear in $\hat a,\hat a^{\dagger}$ and leaves
\begin{equation}
  \begin{split}
  \tilde H=\;&\delta\hat a^{\dagger}\hat a
     +\mu\sum_j|r_2\rangle\langle r_2|_j\\
     &+g\sum_j\!\Big[(\hat a+\alpha)\sigma_j^{+}
                    +(\hat a^{\dagger}+\alpha^{*})\sigma_j^{-}\Big].
  \end{split}
  \label{eq:Htilde}
\end{equation}
In the dispersive regime $|\Delta|\gg g\sqrt{\bar n+1}$ the state $|r_1\rangle$
is adiabatically eliminated. The relevant energy denominator is the separation
between $|r_2,k\rangle$ and $|r_1,k+1\rangle$, which in the rotating frame is
$\mu+\delta k-\delta(k+1)=\Delta$.
\subsection{Exact solution and closure conditions}
Retaining the force term of Eq.~\eqref{eq:Heff_cz}, the evolution is
\begin{equation}
  U(t)=\exp\!\big[\hat n(\beta\hat a^{\dagger}-\beta^{*}\hat a)\big]
       \exp\!\big[-i\Theta(t)\hat n^{2}\big],
  \label{eq:Ufactor}
\end{equation}
with the phase-space trajectory and the enclosed area
\begin{align}
  \beta(t)   &=-i\!\int_0^{t}\!\zeta(s)\,e^{i\delta s}\,ds,
  \label{eq:beta}\\
  \Theta(t)  &=\mathrm{Im}\!\int_0^{t}\!\!ds\!\int_0^{s}\!\!ds'\;
               \zeta(s)\zeta^{*}(s')\,e^{i\delta(s-s')} .
  \label{eq:Theta}
\end{align}
Sector $n$ precesses at $\omega_n=\delta+n\chi$, so loop closure
$\beta_n(t_g)=0$ requires the drive spectrum to vanish simultaneously at the
frequency $\{\delta+n\chi\}$,
\begin{equation}
  \int_0^{t_g}\!\eta(t)\,e^{i(\delta+n\chi)t}\,e^{\kappa t/2}\,dt=0,
  \qquad n=0,1,2 .
  \label{eq:C1}
\end{equation}
For a single loop of constant force $\delta t_g=2\pi$, requiring
$|\chi|/\delta=k_w$, a shaped drive satisfies this condition more generally.  In the constant-force case Eq.~\eqref{eq:Theta}
evaluates to $\Theta=2\pi|\zeta_0/\delta|^{2}$, so that $\Theta=\pi/2$ fixes
\begin{equation}
  |\zeta_0|=\delta/2,
  \label{eq:gatecondition}
\end{equation}
Equation~\eqref{eq:gatecondition} is a leading-order dispersive result.  Since
$\Theta$ scales quadratically with the drive amplitude $A$, so that
$\delta\Theta/\Theta=2\,\delta A/A$, and since the finite-$r$ corrections below
displace $\Theta$ from its ideal value, the amplitude must be calibrated against
the full model.
 
\subsection{Photon-number dependence of the dispersive shift}
 
Diagonalising the block $\{|r_2,k\rangle,|r_1,k+1\rangle\}$, whose coupling is
$g\sqrt{k+1}$ and whose detuning is $\Delta$, gives the exact dressed shift
\begin{equation}
  \delta E(k)=\chi(k+1)-\frac{\chi^{2}}{\Delta}(k+1)^{2}+\dots
  \label{eq:JCexact}
\end{equation}
The quadratic term makes the
effective shift photon-number dependent,
$\chi_{\mathrm{eff}}(k)\simeq\chi[1-(2k+1)\chi/\Delta]$, so that $X$ acquires a
$k$ dependence.  Writing the closure phase as $\varphi(k)=Xk+Yk^{2}$, a phase
linear in $k$ is made trivial, whereas the quadratic coefficient is fixed $-2\pi\frac{\chi}{\Delta}=-2\pi r^{2}$
Since the coefficient is neither zero nor a multiple of $2\pi$, different Fock components
close their loops with different phases, and averaging over the photon-number
distribution dephases the gate by an amount proportional to $\mathcal{O}[r^{4}\mathrm{Var}((k+1)^2)]$.
\section*{Appendix C: bichromatic Raman gate}
\setcounter{subsection}{0}
\subsection{Rotating frame}
After making the optical rotating-wave approximation, the two laser
couplings may be written as
\begin{equation}
\begin{split}
    H_{\mathrm{opt}}(t)
    =\frac{1}{2}\sum_j \Big[
    &\Omega_1(t)e^{-i(\omega_{L1}t+\phi_1)}
        \sigma_j^{r_1,1}
    \\
    &+\Omega_2(t)e^{-i(\omega_{L2}t+\phi_2)}
        \sigma_j^{r_2,1}
    +\mathrm{H.c.}\Big].
    \label{app_raman_labdrive}
\end{split}
\end{equation}
We use the rotating-frame transformation
\begin{equation}
\begin{split}
    \hat U_r(t)=\exp\!\Bigg\{it\Bigg[
    &\omega_{L1}\sum_j\hat n_{r_1,j}
    +\omega_{L2}\sum_j\hat n_{r_2,j}
    \\
    &+(\omega_{L2}-\omega_{L1})
    \hat a^\dagger\hat a\Bigg]\Bigg\}.
    \label{app_raman_frame}
\end{split}
\end{equation}
the Hamiltonian in this frame is
\begin{equation}
    H_r=\hat U_rH\hat U_r^\dagger
    +i\dot{\hat U}_r\hat U_r^\dagger.
    \label{app_raman_frameham_def}
\end{equation}
One optical phase can be absorbed into the definitions of the Rydberg states. With
\begin{equation}
    \phi=\phi_2-\phi_1,
    \label{app_raman_relative_phase}
\end{equation}
The rotating-frame Hamiltonian is then
\begin{equation}
\begin{split}
    H_r={}&\delta_r\hat a^\dagger\hat a
    +\sum_j\Bigg[
    \Delta_1\hat n_{r_1,j}
    +\Delta_2\hat n_{r_2,j}
    +g\left(\sigma_j^+\hat a+\sigma_j^-\hat a^\dagger\right)
    \\
    &\quad+\frac{1}{2}\left(
    \Omega_1(t)\sigma_j^{r_1,1}
    +\Omega_2(t)e^{-i\phi}\sigma_j^{r_2,1}
    +\mathrm{H.c.}\right)\Bigg],
    \label{app_raman_rotating_hamiltonian}
\end{split}
\end{equation}
where
\begin{equation}
\begin{aligned}
    \Delta_1&=\omega_{r_1}-\omega_{L1},
    \qquad
    \Delta_2=\omega_{r_2}-\omega_{L2},
    \\
    \delta_r&=\omega_c-(\omega_{L2}-\omega_{L1}).
    \label{app_raman_detunings}
\end{aligned}
\end{equation}
The atom--cavity detuning is
\begin{equation}
    \Delta=\omega_{r_2r_1}-\omega_c
    =\omega_{r_2}-\omega_{r_1}-\omega_c.
    \label{app_raman_cavity_detuning}
\end{equation}
Equations~\eqref{app_raman_detunings} and
\eqref{app_raman_cavity_detuning} give the exact frame-consistency
condition
\begin{equation}
    \Delta_2-\Delta_1=\Delta+\delta_r.
    \label{app_raman_frame_consistency}
\end{equation}

\subsection{Elimination of the Rydberg manifold}
Let $P = |0\rangle \langle0| +|1\rangle \langle1|$ project onto the logical manifold and let $Q=1-P$ project onto the two Rydberg levels. For one atom in $|1\rangle$, the operator that couples the logical and Rydberg manifolds is
\begin{equation}
\begin{split}
    V_+(t)={}&\frac{1}{2}\left[
    \Omega_1(t)\lvert r_1\rangle
    +\Omega_2(t)e^{-i\phi}\lvert r_2\rangle
    \right]\langle1\rvert,
    \\
    V_-(t)={}&V_+^\dagger(t).
    \label{app_raman_vplus}
\end{split}
\end{equation}
In the basis
$\{\lvert r_1\rangle,\lvert r_2\rangle\}$, the Hamiltonian acting inside the Rydberg manifold is
\begin{equation}
    H_R=
    \begin{pmatrix}
        \Delta_1 & g\hat a^\dagger\\
        g\hat a & \Delta_2
    \end{pmatrix}
    =D+G,
    \label{app_raman_hr}
\end{equation}
with
\begin{equation}
    D=
    \begin{pmatrix}
        \Delta_1 & 0\\
        0 & \Delta_2
    \end{pmatrix},
    \qquad
    G=
    \begin{pmatrix}
        0 & g\hat a^\dagger\\
        g\hat a & 0
    \end{pmatrix}.
    \label{app_raman_dg}
\end{equation}
The Rydberg amplitudes follow their instantaneous stationary values when
\begin{equation}
    \frac{\lvert\Omega_i\rvert}{2\lvert\Delta_i\rvert}\ll1,
    \qquad
    \frac{g\sqrt{k+1}}{\lvert\Delta\rvert}\ll1,
    \qquad
    \frac{\lvert\dot\Omega_i\rvert}
         {\lvert\Delta_i\Omega_i\rvert}\ll1
    \label{app_raman_validity}
\end{equation}
over every appreciably occupied cavity sector $k$. We define the generator to be anti hermitian $S = H_R^{-1}V_+-V_-H_R^{-1}$, and make the transformation $\tilde H = e^{S}He^{-S}$. The projected Hamiltonian is
\begin{equation}
    H_{\mathrm{eff}}
    =\delta_r\hat a^\dagger\hat a
    -\sum_j V_{-,j}H_R^{-1}V_{+,j}.
    \label{app_raman_projected}
\end{equation}
The inverse of $H_R$ can be expanded as
\begin{equation}
\begin{split}
    H_R^{-1}
    &=(D+G)^{-1}
    \\
    &=D^{-1}-D^{-1}GD^{-1}
    +D^{-1}GD^{-1}GD^{-1}+\cdots.
    \label{app_raman_inverse}
\end{split}
\end{equation}
The first term gives the differential optical light shift
\begin{equation}
    -V_-D^{-1}V_+
    =\epsilon(t)\lvert1\rangle\langle1\rvert,
    \label{app_raman_lightshift_step}
\end{equation}
where
\begin{equation}
    \epsilon(t)
    =-\frac{\lvert\Omega_1(t)\rvert^2}{4\Delta_1}
    -\frac{\lvert\Omega_2(t)\rvert^2}{4\Delta_2}.
    \label{app_raman_lightshift}
\end{equation}
The term containing one power of $G$ is third order in the original
couplings and gives
\begin{equation}
\begin{split}
    V_-D^{-1}GD^{-1}V_+
    =\lvert1\rangle\langle1\rvert\Bigg[
    &\frac{g\Omega_1^*(t)\Omega_2(t)}{4\Delta_1\Delta_2}
       e^{-i\phi}\hat a^\dagger
    \\
    &+\frac{g\Omega_1(t)\Omega_2^*(t)}{4\Delta_1\Delta_2}
       e^{i\phi}\hat a\Bigg].
    \label{app_raman_force_step}
\end{split}
\end{equation}
Defining
\begin{equation}
    \lambda(t)=
    \frac{g\Omega_1(t)\Omega_2^*(t)}{4\Delta_1\Delta_2},
    \label{app_raman_lambda_complex}
\end{equation}
and summing over the atoms gives
\begin{equation}
\begin{split}
    H_{\mathrm{eff}}(t)
    ={}&\delta_r\hat a^\dagger\hat a
    +\hat n_1\left[
       \lambda(t)e^{i\phi}\hat a
       +\lambda^*(t)e^{-i\phi}\hat a^\dagger
       \right]
    \\
    &+\epsilon(t)\hat n_1.
    \label{app_raman_effective_full}
\end{split}
\end{equation}
\subsection{Exact evolution under the force Hamiltonian}
It is useful first to retain arbitrary complex pulse envelopes. We define
\begin{equation}
    \xi(t)=\lambda(t)e^{i\phi}
    \label{app_raman_zeta}
\end{equation}
The effective Hamiltonian is
\begin{equation}
    H_{\mathrm{eff}}(t)
    =\delta_r\hat a^\dagger\hat a
    +\hat n_1\left[\zeta(t)\hat a+\xi^*(t)\hat a^\dagger\right]
    +\epsilon(t)\hat n_1.
    \label{app_raman_general_force}
\end{equation}
In the interaction picture of
$\delta_r\hat a^\dagger\hat a$, this becomes
\begin{equation}
    H_I(t)=\hat n_1\left[
    \xi(t)e^{-i\delta_rt}\hat a
    +\xi^*(t)e^{i\delta_rt}\hat a^\dagger
    \right]
    +\epsilon(t)\hat n_1.
    \label{app_raman_interaction_force}
\end{equation}
The commutator at two different times is proportional to
$\hat n_1^2$ and contains no cavity operator. All Magnus terms above
second order therefore vanish. The interaction-picture propagator is
exactly
\begin{equation}
    U_I(t)=
    D\!\left[\beta(t)\hat n_1\right]
    \exp\!\left[i\Theta(t)\hat n_1^2\right]
    \exp\!\left[-i\Phi(t)\hat n_1\right],
    \label{app_raman_magnus_solution}
\end{equation}
where
\begin{equation}
    \beta(t)=-i\int_0^t
    \xi^*(s)e^{i\delta_rs}\,ds,
    \label{app_raman_beta_general}
\end{equation}
\begin{equation}
\begin{split}
    \Theta(t)=\operatorname{Im}
    \int_0^t ds\int_0^s ds'\,
    &\xi^*(s)\xi(s')
    \\
    &\times e^{i\delta_r(s-s')},
    \label{app_raman_theta_general}
\end{split}
\end{equation}
and
\begin{equation}
    \Phi(t)=\int_0^t\epsilon(s)\,ds.
    \label{app_raman_phi_general}
\end{equation}
The cavity disentangles from the atoms whenever
\begin{equation}
    \beta(t_g)=0.
    \label{app_raman_closure_general}
\end{equation}
At closure the remaining cavity rotation is independent of $\hat n_1$, so it cannot carry
logical-state information. For the single-loop gate we take $\lambda$ and $\epsilon$ to be
constant. In a sector containing $n$ atoms in $\lvert1\rangle$, the
Hamiltonian is
\begin{equation}
    H_n=\delta_r\hat a^\dagger\hat a
    +n\left(\xi\hat a+\xi^*\hat a^\dagger\right)
    +n\epsilon.
    \label{app_raman_sector_hamiltonian}
\end{equation}
Using
$D^\dagger(\alpha)\hat aD(\alpha)=\hat a+\alpha$ and choosing
\begin{equation}
    \alpha_n=-\frac{n\xi^*}{\delta_r}
    \label{app_raman_alpha}
\end{equation}
gives
\begin{equation}
    D^\dagger(\alpha_n)H_nD(\alpha_n)
    =\delta_r\hat a^\dagger\hat a
    -\frac{n^2\lvert\xi\rvert^2}{\delta_r}
    +n\epsilon.
    \label{app_raman_displaced_hamiltonian}
\end{equation}
The corresponding cavity displacement is
\begin{equation}
    \beta_n^{(S)}(t)
    =-\frac{n\xi^*}{\delta_r}
      \left(1-e^{-i\delta_rt}\right).
    \label{app_raman_beta_constant}
\end{equation}
The phase enclosed by the trajectory is
\begin{equation}
    \Theta_n(t)
    =\frac{n^2\lvert\lambda\rvert^2}{\delta_r^2}
      \left(\delta_rt-\sin\delta_rt\right).
    \label{app_raman_theta_constant}
\end{equation}
For $\delta_r>0$, one complete loop has duration
\begin{equation}
    t_g=\frac{2\pi}{\delta_r}.
    \label{app_raman_gate_time}
\end{equation}
At this time $\beta_n^{(S)}(t_g)=0$ for every $n$ and the propagator
factorises as
\begin{equation}
    U(t_g)=
    \exp\!\left[
       2\pi i\left\lvert\frac{\lambda}{\delta_r}\right\rvert^2
       \hat n_1^2
    \right]
    \exp\!\left[-i\epsilon t_g\hat n_1\right]
    \otimes\mathbb{1}_{\mathrm{cav}}.
    \label{app_raman_closed_propagator}
\end{equation}
For negative $\delta_r$, the geometric phase changes sign and the same
construction follows after reversing the orientation of the loop.

\subsection{Controlled-phase condition}

For two atoms we write
\begin{equation}
    \hat n_1=\hat n_{1,1}+\hat n_{1,2},
    \qquad
    \hat n_1^2=\hat n_1+2\hat n_{1,1}\hat n_{1,2}.
    \label{app_raman_nsquare}
\end{equation}
The term proportional to $\hat n_1$ is local. After correcting this term
and the light-shift phase, the nonlocal propagator is
\begin{equation}
    U_{\mathrm{nl}}=
    \exp\!\left[2i\theta
    \hat n_{1,1}\hat n_{1,2}\right],
    \qquad
    \theta=2\pi
    \left\lvert\frac{\lambda}{\delta_r}\right\rvert^2.
    \label{app_raman_nonlocal}
\end{equation}
A CZ gate requires $2\theta=\pi$, the shortest solution is $\theta=\frac{\pi}{2}$, $t_g=\frac{\pi}{|\lambda|}$. The uncorrected logical operation is
$\{|00\rangle,|01\rangle,
|10\rangle,|11\rangle\}$ is
\begin{equation}
    U(t_g)=\operatorname{diag}(1,i,i,1).
    \label{app_raman_diag_gate}
\end{equation}
which is locally equivalent to the CZ gate
\subsection{Balanced operating point}
The virtual Rydberg amplitudes of an atom initially in
$\lvert1\rangle$ are, to leading order,
\begin{equation}
    c_{r_1}=-\frac{\Omega_1}{2\Delta_1}c_1,
    \qquad
    c_{r_2}=-\frac{\Omega_2e^{-i\phi}}{2\Delta_2}c_1.
    \label{app_raman_virtual_amplitudes}
\end{equation}
The corresponding Rydberg population is
\begin{equation}
    p_R=\frac{|\Omega_1|^2}{4\Delta_1^2}
    +\frac{|\Omega_2|^2}{4\Delta_2^2}.
    \label{app_raman_pr_general}
\end{equation}
The dispersive cavity shift of the two Rydberg levels is
\begin{equation}
    H_{\mathrm{disp}}
    =\frac{g^2}{\Delta}\left[
    (\hat a^\dagger\hat a+1)\hat n_{r_2}
    -\hat a^\dagger\hat a\hat n_{r_1}
    \right].
    \label{app_raman_rydberg_dispersion}
\end{equation}
Substitution of Eq.~\eqref{app_raman_virtual_amplitudes} into
Eq.~\eqref{app_raman_rydberg_dispersion} gives the leading
cross-Kerr term
\begin{equation}
    H_{\mathrm{Kerr}}^{(4)}
    =\chi_R\hat n_1\hat a^\dagger\hat a,
    \label{app_raman_kerr}
\end{equation}
where
\begin{equation}
    \chi_R=\frac{g^2}{4\Delta}
    \left(
    \frac{|\Omega_2|^2}{\Delta_2^2}
    -\frac{|\Omega_1|^2}{\Delta_1^2}
    \right).
    \label{app_raman_chir}
\end{equation}
The leading cross-Kerr term vanishes when the two optical tones admix
equal Rydberg populations,
\begin{equation}
    \frac{|\Omega_1|}{|\Delta_1|}
    =
    \frac{|\Omega_2|}{|\Delta_2|}.
    \label{app_raman_equal_admixture}
\end{equation}
The differential light shift vanishes when
\begin{equation}
    \frac{|\Omega_1|^2}{\Delta_1}
    +
    \frac{|\Omega_2|^2}{\Delta_2}=0.
    \label{app_raman_light_cancel}
\end{equation}
Together with Eq.~\eqref{app_raman_frame_consistency}, these two
conditions select
\begin{equation}
    |\Omega_1|=|\Omega_2|=\Omega,
    \qquad
    \Delta_1=-\frac{\Delta+\delta_r}{2},
    \qquad
    \Delta_2=\frac{\Delta+\delta_r}{2}.
    \label{app_raman_balanced_exact}
\end{equation}
When $|\delta_r|\ll|\Delta|$, this reduces to
\begin{equation}
    \Delta_2\simeq-\Delta_1\simeq\frac{\Delta}{2}.
    \label{app_raman_balanced_approx}
\end{equation}
At the balanced point,
\begin{equation}
    |\lambda|
    =\frac{g\Omega^2}{(\Delta+\delta_r)^2}
    \simeq\frac{g\Omega^2}{\Delta^2},
    \label{app_raman_lambda_balanced}
\end{equation}
and
\begin{equation}
    p_R=\frac{2\Omega^2}{(\Delta+\delta_r)^2}
    =\frac{2|\lambda|}{g}.
    \label{app_raman_pr_balanced}
\end{equation}
The Rydberg
population of one atom initially in $|1\rangle$ is $ p_Rt_g=2\pi/{g}$. The mean value of $\hat n_1$ over the four computational basis states is
1. For equal Rydberg decay rates $\Gamma_r$, the loss is therefore
\begin{equation}
    \varepsilon_{\mathrm{Ryd}}
    \simeq\Gamma_rp_Rt_g
    =\frac{2\pi\Gamma_r}{g}.
    \label{app_raman_decay_floor}
\end{equation}
\subsection{Cavity loss and residual thermal sensitivity}
At the CZ operating point, the separation between the phase-space
trajectories belonging to sectors $n$ and $m$ is
\begin{equation}
    |\beta_n^{(S)}(t)-\beta_m^{(S)}(t)\rvert^2
    =\frac{(n-m)^2}{4}
    \left|1-e^{-i\delta_rt}\right|^2.
    \label{app_raman_path_separation}
\end{equation}
The integral over one loop is
\begin{equation}
    \int_0^{t_g}
    |-e^{-i\delta_rt}|^2dt
    =2t_g.
    \label{app_raman_path_integral}
\end{equation}
Thermal photons in the cavity multiplies $n$ and $m$ by
\begin{equation}
    C_{nm}=\exp\!\left[
    -\frac{\kappa(2\bar n+1)t_g}{4}(n-m)^2
    \right].
    \label{app_raman_coherence_factor}
\end{equation}
For the two-qubit sector labels $\{0,1,1,2\}$, expansion of
Eq.~\eqref{app_raman_coherence_factor} to first order in $\kappa t_g$ gives
\begin{equation}
    1-\overline F_{\mathrm{cav}}
    \simeq\frac{\kappa(2\bar n+1)t_g}{5}.
    \label{app_raman_cavity_average_error}
\end{equation}
At the balanced point the contribution in
Eq.~\eqref{app_raman_kerr} vanishes. The remaining photon-number
dependence enters through higher-order corrections. If the relative
force correction obeys
\begin{equation}
    \frac{\delta\lambda}{\lambda}
    =O\!\left(r^2k\right),
    \qquad
    r=\frac{g}{\Delta},
    \label{app_raman_force_correction}
\end{equation}
then the corresponding phase error is linear in $r^2k$ and the
infidelity is quadratic. Averaging over a thermal distribution gives
\begin{equation}
    \varepsilon_{\mathrm{th}}
    =O\!\left[r^4\operatorname{Var}(k)\right]
    =O\!\left[r^4\bar n(\bar n+1)\right].
    \label{app_raman_thermal_residual}
\end{equation}
The leading analytic error budget is therefore
\begin{equation}
\begin{split}
    1- F\simeq{}
    &\frac{2\pi\Gamma_r}{g}
    +\frac{\kappa(2\bar n+1)t_g}{5}
    +O\!\left[r^4\bar n(\bar n+1)\right].
    \label{app_raman_error_budget}
\end{split}
\end{equation}
\end{document}